# Temperature Dependent Na-ion Conduction and its Pathways in the Crystal Structure of the Layered Battery Material $Na_2Ni_2TeO_6$

A. K. Bera[1] and S. M. Yusuf[1,2*]

[1]*Solid State Physics Division, Bhabha Atomic Research Centre, Mumbai 400085, India*

[2]*Homi Bhabha National Institute, Anushaktinagar, Mumbai 400094, India*

## Abstract

Na-ion conduction, and correlations between Na-ion conduction pathways and crystal structure have been investigated as a function of temperature in the layered battery material $Na_2Ni_2TeO_6$ by impedance spectroscopy and neutron diffraction, respectively. The impedance data reveal an ionic conductivity $\sigma \approx 2\times10^{-4}$ S/m at 323 K which strongly enhances with increasing temperature and attains a high value of $\sim$ 0.03 S/m at 423 K. The temperature dependent conductivity data show an Arrhenius type behavior with average activation energy ($E_a$) of $\sim$ 0.58(3) eV for $T \geq 383$ K. By employing soft-bond valence sum analysis of the neutron diffraction patterns we experimentally demonstrate the site-specific Na-ion conductions, through visualization of microscopic sodium-ion conduction pathways, and verify the recent theoretical results of Molecular dynamic simulation. Our results reveal two-dimensional Na-ion conduction pathways that are confined within the *ab* planes of Na layers. Crystal structural study indicates that the layered structure involving Na ion layers is responsible for high ionic conductivity, and the local crystallographic environment of Na ion sites is responsible for site-specific conductivity. Our study further reveals that up to 500 K, the ionic conduction is governed by the Na ions located at the Na1 and Na2 sites, whereas, all the Na ions located at the three Na-sites contribute to the conduction process above 500 K. Our neutron diffraction study also establishes that the crystal structure of $Na_2Ni_2TeO_6$ is stable at least up to 725 K (the highest measured temperature), however, with an anisotropic thermal expansion ($\alpha_c/\alpha_a \sim 3$).

---

*email: smyusuf@barc.gov.in, Phone: +91 22 25595608

I. INTRODUCTION

The search for efficient ionic conductors for high-energy-density battery applications under renewable energy technology (to replace conventional fossil fuels) is one of the most active fields in materials research [1]. A significant advancement in the development of high energy-density Li-ion batteries was witnessed over the last decades. Li-ion batteries are favorite for small scale applications and have been leading the present market. Lithium is, however, expensive as the availability of Li is rather limited. Alternatively, sodium conducting materials are more promising for energy storage applications, as Na is naturally abundant, second lightest alkali metal after lithium and it has a suitable redox potential, thus Na is a better candidate for medium and large scale energy storage applications[2-3].

One of the major challenges in Na-ion batteries is to design improved materials for better efficiency. In this regard, the family of Na based compounds $Na_2X_2TeO_6$ (where $X$ = Co, Ni, Zn, Mg, and Fe) has drawn good attention in recent years[4-8]. These compounds are reported to have significant Na conductivity at room temperature, suitable for application in a Na-ion battery [4]. Among these compounds, the compound $Na_2Ni_2TeO_6$ was reported to show the highest ionic conductivity (~ 0.0008-0.0034 S/m at room temperature) [6, 9] which is comparable to the value reported for the $\beta$-alumina, a well-known material for solid electrolyte [10]. Moreover, a highly reversible Na extraction/insertion process was reported for $Na_2Ni_2TeO_6$ with a reasonable capacity of 100 mAh $g^{-1}$ in the 4.35 to 3.00 V voltage range [11]. Recent molecular dynamics simulations[9, 12] on $Na_2Ni_2TeO_6$ reported that the Na-ion conductions pathways were confined within the Na-ion layers in the *ab* plane, and such 2D pathways were highly anisotropic and Na-ion site specific. However, the experimental verification of such anisotropic Na-ion conduction pathways is lacking. In the present study, we provide an experimental verification of such anisotropic Na-conduction pathways by a comprehensive neutron diffraction study on $Na_2Ni_2TeO_6$. Such an understanding of the microscopic mechanism of ionic conduction is essential to control and improve ionic transport in solid materials. We also provide the temperature dependent impedance spectroscopy data over 323 to 423 K revealing its ionic conductivity.

Here, we report a detailed experimental investigation of the crystal structure of Na$_2$Ni$_2$TeO$_6$ over a wide temperature range of 300-725 K (covering the possible operating range for its battery application) by neutron powder diffraction. We report the visualization of the sodium-ion conduction pathways through high-temperature neutron powder diffraction experiments. We have employed the soft-bond valence sum analysis of the measured experimental neutron diffraction patterns to determine Na-ion conduction pathways and the available space for Na-ion conduction within the unit cell, one of the main decisive parameters for the Na-ion conductivity. Soft-bond-valence is an efficient method to determine accessible sites for mobile ions that are obtained by analyzing the valence mismatch of moving ions[13-18]. Such low energy Na-ion conduction channels within the crystal structure are determined by the bond valence energy landscape (BVEL) map. The determined Na-ion conduction pathways are found to be two-dimension and confined within the Na layers in the *ab* planes. The intermediate metal-oxide layers, formed by edge-sharing (Te/Ni)O$_6$ octahedra, are found to be compact and impenetrable for Na ions. We also reveal the temperature dependence of such Na-ion conduction pathways over 300-725 K and their role on the Na-ion conductivity, measured by an impedance spectroscopy. Our comprehensive investigation provides the understanding of the conduction pathways in the crystalline structure of Na$_2$Ni$_2$TeO$_6$ facilitating ionic conduction in the solid state material.

## II. EXPERIMENTS

Polycrystalline samples of Na$_2$Ni$_2$TeO$_6$ were synthesized by solid state reaction method. A stoichiometric mixture of Na$_2$CO$_3$ (99.9 %), NiO (99.99 %) and TeO$_2$ (99.99 %) was heated at 950 °C in air for a total period of 72 hrs. The mixture was powdered and then pelletized intermediately after every heating module of 24 hrs. The quality of the samples was checked by the laboratory x-ray diffraction.

Impedance measurements were performed in the frequency (ν) range 1–10$^6$ Hz with an ac-electric field amplitude of 0.2 V by using Newtons4th Ltd, UK make PSM1735 – NumetriQ impedance analyzer over the temperature range of 323 –423 K with a step size of 5 K. For the impedance measurements, disc-shaped pellets (diameter = 10 mm, thickness ∼ 0.6 mm) were prepared at room temperature and then annealed at 900 °C to increase the density. The density of the pellets was determined to be ∼ 90% of the theoretical value, and the measured impedance data were accordingly corrected. Silver paste was uniformly coated on both sides of the pellets and dried before the recording of the impedance spectra. *EIS Spectrum Analyser* software was used for the impedance data analysis[19].

Neutron powder diffraction measurements were performed by using the neutron powder diffractometer-II ($\lambda$ = 1.2443 Å) at the Dhruva reactor, Bhabha Atomic Research Centre, Mumbai, India over 300-725 K. For these measurements, a furnace was used. The powder sample was filled inside a quartz tube which was placed inside a vanadium can for the neutron diffraction measurements. The contribution from the sample environment (quartz tube, vanadium can, and the furnace) was estimated separately by an empty sample can measurement. The experimental neutron diffraction patterns were analyzed by Rietveld refinement method by using the FULLPROF computer program[20]. The soft-bond valence sum analysis was performed by using the Bond_STR program[21] available in the Fullprof suite.

## III.    RESULTS AND DISCUSSION

We first present the results of the impedance spectroscopy study on $Na_2Ni_2TeO_6$. The frequency-dependent real (Z′) and imaginary (-Z′′) parts of impedance at selected temperatures of $Na_2Ni_2TeO_6$ are shown in Figs. 1(a) and 1(b), respectively. At 323 K, with increasing frequency (*f*), the Z′ value shows an almost constant value (∼ 85 kOhm) up to *f* ∼ 200 Hz and then shows a rapid decrease (frequency activated) with a further increase of the frequency up to 1 MHz. With increasing temperature, the nature of the frequency dependence also changes. On the other hand, the -Z′′ vs *f* curves show distinguishing features; a decay at low frequencies and two peaks centered around ∼ 0.5 and 50 kHz. With increasing temperature, the peak positions shift towards higher frequencies, and simultaneously the peak becomes broad with a decrease in its peak height. The broadening of the peaks suggests a spreading of relaxation time and hence, the existence of a temperature dependent relaxation phenomenon in the present compound. It further suggests that the time relaxation is of non-Debye type. With increasing temperature, an enhancement of the −Z″-value over low frequencies is also evident.

The Nyquist plots or Cole–Cole plots (−Z″vs Z′) are shown in Fig. 1(c) for few selected temperatures in order to estimate the contributions of electrodes, grain, and grain boundaries. Each of the −Z″ vs Z′ curves consists of two partially overlapping depressed semi-circles and a sharp linear increase at higher Z′ values (low frequencies). This type of behavior is an indication of ionic conduction with blocking electrodes[22]. Evsteigneeva *et al.* [4] reported that the conductivity in the studied compound $Na_2Ni_2TeO_6$ is essentially pure ionic in nature with a negligible electronic contribution. The two semi-circles in our data appear due to the contributions from grain interior ($g_{int}$) and grain boundaries ($g_b$) of polycrystalline sample[23]. The low-frequency impedance can be attributed to the electrode-sample interface[23]. With increasing temperature, the relative contributions of the semicircles decrease monotonically and shift towards higher frequencies. On the other hand, the relative contribution of the low-frequency linear response increases with increasing temperature. The depressed semicircles are due to distributed microscopic material properties and can be modelled by the constant phase element (CPE). The equivalent circuit for the observed curves (two depressed semicircles and an inclined straight line) can be accounted by a series of (i) parallel combination of the CPE1 and resistance R1 for grain interior; (ii) parallel combination of CPE2 and resistance R2 for grain boundary; and (iii) the CPE3 of the interface [inset of Fig. 1(c)]. The −Z″ vs Z curves were fitted with the equivalent circuit and the individual contributions of grain interior, and grain boundaries and electrodes were estimated. The fitted curves as per the equivalent circuit are shown by the solid lines in Fig. 1(c). The derived conductivity values are plotted as a function of temperature in Fig. 1(d) as the Arrhenius plot. The Arrhenius equation, utilized for the ionic conduction data analysis, is given by σ = ($A/T$) exp(−$E$a/$RT$), where $E$a is the activation energy for ionic conduction, and $R$ is the universal gas constant. A is the proportionality constant. The Arrhenius plot [Fig. 1(d)] shows a curvature over the measured temperature range. Such curvature in the Arrhenius plot was reported in literature for various systems, and the deviation from the linear behaviour was explained based on several physical processes, such as tunnelling or hopping of charge carrier across the lattice sites [24], electronic spatial potential fluctuations due to the structural variations [25], intrinsic and extrinsic regions of ionic conductors [26], and hydration process [27]. For the present compound, the deviation of the Arrhenius plot from a linear behavior can be related to the observed anisotropic crystallographic structural changes with temperature (Fig. 4). We have estimated the values of activation energy ($E_a$) by fitting the measured experimental data over the higher ($T$ > 383 K) as well as lower ($T$ < 353 K) temperature regions with two straight lines. The fittings yield $E_a$ =0.39 eV (for $T$ < 353 K), and $E_a$ =0.58 eV (for $T$ > 383 K). The derived value of activation energy $E_a$ =0.58 eV for $T$ > 383 K is in good agreement with the literature value of 0.53 eV for $Na_2Ni_2TeO_6$, as reported by Evstigneeva, *et al.*, [4] by fitting the impedance data over the temperature range of 323-573 K.

We present below the neutron diffraction results to gain an insight into the microscopic $Na^+$-ion conduction mechanism and its pathways. The Rietveld refined neutron diffraction patterns recorded at 300 and 725 K are shown in Fig 2. The analysis reveals that the compound crystallizes in the hexagonal symmetry with space group *P*6$_3$/*mcm* (No. 193) at both 300 and 725 K. Hence, no crystal structural transition is evident over the studied temperature range. The refined values of lattice parameters, atomic positions, and isotropic thermal parameters at 300 and 725 K are given in Table **I**. The hexagonal crystal structure of $Na_2Ni_2TeO_6$ consists of alternating layers of $(Ni/Te)O_6$ octahedra and Na ions along the crystallographic *c* axis (Fig. 3). Within the $(Ni/Te)O_6$ octahedral layers, edge shared $NiO_6$ octahedra form regular honeycomb lattice, with $TeO_6$ octahedron being at the center of the honeycomb lattice. These layers are well separated, by a distance of ~ 3.48 Å, along the *c* axis by an intermediate Na layer. The intermediate Na-layers are formed by $NaO_6$ trigonal prisms sharing their rectangular faces with the adjacent Ni/Te-metal oxide layers; thus providing wide passages for ionic transport. Within the intermediate Na-layers, there are three Na-sites in the Na layer and all of them are partially occupied. The site occupancies of the Na1(6*g*), Na2(4*c*), and Na3(2*a*) sites are derived to be 0.37(2), 0.18(3) and 0.11(1), respectively, revealing a lot of vacancies in the Na-sites (as the values are significantly less than 1) and the values are in good agreement with the previous reports [4, 7]. Therefore, among the total Na-ions ~70, 23; and 7 wt% are located at Na1, Na2 and Na3 sites, respectively. The total amount of the Na ions, estimated from the refined values of occupation numbers, is ~ 1.6. Therefore, the most interesting features of the crystal structure are: (i) layered type crystal structure formed with purely Na layers which are well separated from the metal oxide layers of (Te/Ni)$O_6$ octahedra, (ii) there are three crystallographic sites for Na ions all of which are partial occupied by Na ions, and (iii) a disordered distribution of Na ions among the three crystallographic sites. Such features facilitate high ionic conductivities as found from impedance spectroscopy study (Fig. 1). Moreover, for all the three Na-sites, Na ions are located within the Na$O_6$ trigonal biprismatic environment formed by six oxygen ions. Among these six oxygen ions, three oxygen ions are from the top Ni/Te$O_6$ layer (along the c axis), and the rest of the three oxygen ions are from the bottom Ni/Te$O_6$ layer [Fig. 3(b)-(d)]. For the Na1 site, three oxygen ions on the top/bottom layer are from three different octahedra (two from two Ni$O_6$ and one from Te$O_6$), respectively [Fig. 3(b)]. On the other hand, for the Na2 and Na3 sites, all three oxygen ions at the top or bottom layers are from a single octahedron of Ni$O_6$ and Te$O_6$, respectively [Fig. 3(c)-(d)]. Within a given *ab* plane, the Na1 site has two neighbors of the Na2 sites ($d_{Na1-Na2}$ = 1.623 Å) and one neighbor of the Na3 site ($d_{Na1-Na3}$ =1.992 Å). On the other hand, for both the Na2 and Na3 sites, all three neighbors are the Na1 sites alone with $d_{Na1-Na2}$ = 1.623 Å and $d_{Na1-Na3}$ =1.992 Å, respectively. Such local site-specific crystal structural environments have a significant role in the site-specific Na-ion conduction as discussed in next section.

The temperature dependent neutron diffraction patterns are shown in Fig. 4(a). With increasing temperature, all the Bragg peaks shift towards lower scattering angle revealing an increase of lattice constants and the unit cell volume. A separation of the Bragg peaks at 2θ = 49 degrees is also evident, indicating an asymmetric thermal expansion of the lattice. The detailed temperature evolutions of few selected Bragg peaks are shown in Fig. 4(b). A larger shift is found for the Bragg peaks having higher *l*-values as compared to the Bragg peaks those are having *l* = 0 or with lower *l*-values. This reveals a stronger temperature dependence of the lattice constant *c* than that of the lattice constants *a* and *b*. The temperature dependences of the lattice constants (*a* and *c*) and unit cell volume (*V*) are shown in Fig. 5(a) and Fig. 5(b), respectively, over 300–725 K. The relative changes of the *a*, *c* and *V* with respect to their values at 300 K as a function of temperature are shown in Fig. 5(c). Our study reveals that the compound remains structurally stable over the studied temperature range (300-725 K) with a small change (~2.7%) in the unit cell volume $V$. However, an asymmetric thermal expansion is evident, where a negligible change ($\sim$ 0.6%) is found along the $a/b$ axes, while the change in $c$ value is $\sim$ 1.6%. The thermal-expansion coefficients, obtained from the linear fitting, are $\alpha_a$ = 1.299×10$^{-5}$ K$^{-1}$ along the $a$ axis, $\alpha_c$ = 3.858×10$^{-5}$ K$^{-1}$ along the $c$ axis, and $\alpha_V$ = 6.507 × 10$^{-5}$ K$^{-1}$ for the unit cell volume. The $\alpha_c$ is $\sim$ 2.97 times higher than the $\alpha_a$ indicating a strongly anisotropic thermal expansion of the lattice. Such an asymmetric thermal expansion of lattice parameters was also reported by Karna et al. [7]. It may be mentioned here that the temperature dependent lattice parameters show a weak anomaly at $\sim$ 500 K which may indicate some minor crystal structural change. A comprehensive high-resolution neutron diffraction measurements around this temperature with finer temperature steps along with other physical properties measurements are required for a detailed investigation of this anomaly and could be attempted in future. The anisotropic thermal expansion in Na$_2$Ni$_2$TeO$_6$ leads to an increase of the separation between two (Ni/Te)O$_6$ metal oxide layers from 3.48 Å to 3.55 Å with increasing temperature from 300 to 725 K. Such an increase in the separation between the metal oxide layers may also facilitate an enhancement of the Na$^+$ ion conductivity with increasing temperature. This is further evident from our soft-bond-valence sum (BVS) calculation (discussed in next section) that reveals a linear increase of the available space for Na-ion conduction within the unit cell with increasing temperature.

In order to gain insight into the Na$^+$ ion conduction pathways, we have employed a soft bond-valence sum analysis, a new method that is developed recently and successfully applied to estimate the ion-conduction pathways in several fuel cell[18] and battery materials [15, 28-30] from the experimental diffraction patterns. In this calculation, accessible sites for mobile Na-ions are identified by using an empirical relationship between the bond length $R$ and bond valence $S_{Na-O}$,

$S_{Na-O} = exp[(R_0 - R)/b]$, (1)

where $R_0$ and $b$ are empirical constants [31]. The values of $R_0$ (corresponding to the ideal bond length) and $b$ are considered to be 1.803 Å and 0.37 Å (as implemented in the bondSTR program [21]) for the calculations. It is reported that the bond valence parameters ($R_0$ and $b$) attain their constant values for a global cut off distance beyond 6 Å [31]. For the present work, the global cut off distance is considered to be 8 Å at which $R_0$ and $b$ are constants. The values of bond-lengths ($R$) are calculated from the Rietveld refined atomic positions as discussed in the crystal structure section. The BVS, i.e., the sum of the valences over all the six Na-O bonds (in the present case), should be ideally equal to the oxidation state or the valence of the Na ions which is '+1'. Any deviation of the BVS from the ideal value of '+1' indicates the possibility of a positional or dynamic disorder of Na ions, or a lowering of symmetry in the crystal system [7]. The soft-BVS (ΔV) is defined by the absolute difference between the calculated BVS of Na and the ideal value of the valence (i.e., +1 for Na$^+$) as

$$|\Delta V(Na)| = |\sum_{i=1}^{6} S_{Na-O} - V_{ideal}(Na)| \qquad (2).$$

The bond-valence energy landscape (BVEL) maps are obtained from the soft-bond valence sum (BVS) parameters by transforming valence into energy units by using a Morse-type potential and a screened Coulomb potential for the attractive and repulsive parts, respectively [28, 32]. The BVEL is estimated from the experimentally derived ΔV value from neutron diffraction data, at each point on a 3D grid within the unit cell and then plotted as an isosurface using the VESTA software [33]. The Rietveld refined parameters (Table I) are utilized as an input to the soft bond valence method. The BVEL map provides the most probable Na$^+$ ion conduction pathways within the crystalline structure.

The calculated BVEL maps (correspond to the Na-ion conduction pathways) for a bond valence energy $\Delta E$ =0.6 eV, just higher than the activation energy $E_a$ =0.58(3) eV (as determined from conductivity data in Fig. 1), are shown in Figs. 6 and 7. It is apparent from Fig. 6(a) that Na conduction occurs solely within the Na-layers in the *ab* planes. The intervening (Ni/Te)O$_6$ octahedral layers (well separated from the Na layers) do not allow any movement of Na-ions along the *c* axis. Hence, conductivity along the *c* direction is fully suppressed and the movement of Na$^+$ ions is restricted within the two-dimensional (2D) Na-layers. The present study provides an experimental understanding of the 2D conduction pathways for Na$^+$ ions in the present compound as proposed recently by molecular dynamics simulations[9]. For the 2D ion conductions, as no other ions are present within the Na layers, there are no scattering centers due to other ions. This leads to the higher observed ionic conduction in this compound (Fig. 1). It is further evident from Figs. 6(b) and 6(c) (data presented for 300 K) that the conductions of Na$^+$ ions are through Na1 and Na2 sites, in rotary-like pathways within the 2D *ab* plane. Such 2D circular Na$^+$ conduction pathways are in good agreement with that reported by the molecular dynamics simulations [12]. Conductivity study on a single crystal could be helpful to support the 2D nature of ionic conductivity as revealed from the present soft-BVS analysis.

The contribution of individual Na-site to the Na-ion conduction has been estimated by performing BVEL map analysis for various bond valence energies $\Delta E$ over the range =0.1-1 eV. Such BVEL maps for two representative $\Delta E$ = 0.15 and 0.9 eV are shown in Figs. 6(e) and 6(f). It is evident that a higher energy $\Delta E \sim 0.9$ eV is required at 300 K to make a continuous connection of Na-conduction pathways through the Na3 site. The above observations reveal that the Na ions which are located at the Na2 site are most mobile [Fig. 6(e)] where a considerable accessible area for Na-ion migration is available for an energy as low as $\Delta E$ =0.15 eV. The intermediate and least mobile Na ions are situated at the Na1 and Na3 sites, respectively. It may be mentioned here that the individual Na-sites are connected differently to the adjacent Ni/TeO$_6$ layers along the $c$ axis. Here, for the most mobile site Na2, each Na2O$_6$ trigonal biprism is connected to two NiO$_6$ octahedra, one of the top Ni/TeO$_6$ layer and the other one of the bottom Ni/TeO$_6$ layer, respectively [Fig. 3(c)]. For the least-mobile site Na3, each Na3O$_6$ trigonal biprism is connected to two TeO$_6$ octahedra, one of the top Ni/TeO6 layer and the other one of the bottom Ni/TeO6 layer, respectively [Fig. 3(d)]. Whereas, for the intermediate mobile site Na1, each Na1O$_6$ trigonal biprism is connected to three octahedra (one TeO$_6$ and two NiO$_6$) from the top layer and three (one TeO$_6$ and two NiO$_6$) octahedra from the bottom layer [Fig. 3(b)]. Therefore, the individual Na-ion site has a different crystal structural environment that has a role on the Na conduction in the $ab$ plane. Our experimental understanding on the site-specific Na-ion conductivities (discussed above) is in excellent agreement with the results reported from the molecular dynamics calculations [12]. Owing to the low concentration of all of the Na-sites (Table-I), the Na$^+$–Na$^+$ repulsion is expected to be negligible. The total volume fraction of the unit cell available for the Na-ion mobility was estimated by integration over all the pathways within a unit cell. The volume fraction of the unit cell for the Na-ion mobility increases linearly with the bond valance energy $\Delta E$ [Fig. 6(d)].

The temperature dependence of the BVEL maps (Na-conduction pathways) is shown in Figs. 7(a-c) for three representative temperatures $T$ = 300, 500 and 725 K, respectively. With increasing temperature, Na-ion mobility for all the three Na sites increases. Especially the mobility of the Na3 site increases significantly with increasing temperature, and the conduction pathways make a continuous path also through the Na3 site above ∼ 500 K. This facilitates the possibility of Na conduction through Na3 site at higher temperatures. With increasing temperature as the thermal energy of the Na increases, the Na-ions start to explore higher energy regions of the energy landscape and thus, include the Na3 site in the conduction process at $T \geq 500$ K. A linear increase of the volume fraction of the unit cell available for the Na-ion mobility is found with increasing temperature [Fig. 7(d)]. Here we would like to recall that $Na_2Ni_2TeO_6$ shows an asymmetric thermal expansion with $\alpha_c/\alpha_a \sim 3$ (Fig. 5). Therefore, the enhancement of the available volume fraction for the Na-ion mobility may also related to the increase of the lattice constant $c$ with temperature, hence, the separation between the Ni/TeO$_6$ layers. This indicates that an enhancement of the ionic conductivity in the studied compound is possible by increasing the separation between two Ni/TeO$_6$ layers along the $c$ axis by a proper crystal engineering. Thus, the present understanding would be useful to design new battery materials with high Na-ionic conduction.

To get further insight of the Na ion conductions, the thermal displacement parameters of the atoms are estimated from the Rietveld refinements of neutron diffraction patterns. We find that the $B_{iso}$ values for Na-ions enhance strongly with increasing temperature. Whereas, a slight change of the occupation numbers of the Na sites with the increasing temperature is found (Fig. 8). At room temperature, a large displacement parameter has been found for the Na sites ($B_{iso}$= 0.012 Å$^2$) as compared to the other atoms i.e. Ni, Te and O in the crystal structure [Fig. 8(b)] in agreement with the previous reports [4, 7]. Different $B_{iso}$ values for three crystallographic Na1, Na2, and Na3 sites may be expected, however, a trial refinement with individual $B_{iso}$ values resulted into a divergence. Hence, equal $B_{iso}$ value is considered for all the three Na sites during refinements which gives a convergence of the solution. High momentum resolution neutron diffraction data are required to refine the $B_{iso}$ values for individual Na sites. Nevertheless, the average $B_{iso}$ value for the Na sites provides important information as a function of temperature which is discussed below. Our analysis reveals that with increasing temperature, the $B_{iso}$ value for Na sites is strongly enhanced from 0.012 Å$^2$ at 300 K to 0.043 Å$^2$ at 725 K, revealing that the average displacement of the Na ions increases from about 0.11 Å at room temperature to 0.21 Å at 725 K. This divulges that Na ions have shallow potential wells which are responsible for fast and high ionic conduction as established by our ionic conduction study (Fig. 1). Such a strong enhancement of the $B_{iso}$ value of Na ions was reported for the Na-ion battery material $Na_3TiP_3O_9N$ [29] and the phenomenon was correlated to the dynamic displacements of the Na ions. Moreover, the molecular dynamics simulations on the present compound $Na_2Ni_2TeO_6$ [9] reported that the thermal amplitude of the Na ions at 600 K is about 0.2 Å which gives a nice agreement with the value ∼ 0.192 Å at 600 K determined from the derived $B_{iso}$ value from the present neutron diffraction study (Fig. 8). The $B_{iso}$ value (0.037 Å$^2$) at 600 K is much higher than the value (0.012 Å$^2$) at room temperature which reveals a strong enhancement of the $B_{iso}$ value of Na ions with increasing temperature.

Now we shed some light on the applied aspects of the present compound. Previous reports reveal that $Na_2Ni_2TeO_6$ could be used as a cathode material in a Na-ion battery. Electrode (cathode) performance of $Na_2Ni_2TeO_6$ is tested in a coin cell by cyclic voltammetry[11]. Capacity versus voltage curves for the charge/discharge cycle of $Na_2Ni_2TeO_6$ reveal that the sodium de/intercalation reaction is highly reversible for the sodium concentration range ($Na_xNi_2TeO_6$) x∼ 0-1.6[11]. Moreover, the charge/discharge (sodium de/intercalation reaction) profile for $Na_2Ni_2TeO_6$ exhibits two biphasic plateaus at 3.6 and 4.4 V vs $Na^+$/Na which are related to two different $Na^+$ ion orderings at extraction of 1/3 and 2/3 of the sodium ions per formula unit, respectively. A discharge specific capacity of 110 mAhg$^{-1}$ is achieved as compared to the maximum theoretical capacity of 138.5 mAhg$^{-1}$ on the extraction of two Na ions per formula unit to form $Ni_2TeO_6$. The studied compound $Na_2Ni_2TeO_6$ can also have application as an electrolyte material due to its high $Na^+$ ionic conductivity at room temperature with a negligible electronic conductivity[4]. However, the application of $Na_2Ni_2TeO_6$ as an electrolyte is limited due to small electrochemical window over 3.5-4.5 V as well as presence of peaks due to Na-ion orderings at 3.6 and 4.4 V[11].

## IV. CONCLUSIONS

In summary, the temperature dependent crystal structure, $Na^+$-ion conduction pathways, and ionic conductivity of the layered material $Na_2Ni_2TaO_6$ have been investigated by powder neutron diffraction and impedance spectroscopy. Our experimental results reveal that $Na^+$ ions occupy the interlayers of the $Ni/TeO_6$ polyhedral sheets, and such a structural framework facilitates a high ionic conductivity around room temperature. We report the visualization of the sodium-ion conduction pathways in $Na_2Ni_2TeO_6$ through the soft-bond valence sum analysis of the high-temperature neutron powder diffraction patterns. Our comprehensive BVEL map analyses reveal a 2D Na-ionic conduction within the Na-ion layers and provide site-specific contribution of Na-ion conduction. At room temperature only the Na ions located at the Na1 and Na2 sites, among the three Na crystallographic sites, are involved in the ionic conduction process. However, with increasing temperatures above ~ 500 K, continuous ionic conduction pathways occur involving the Na3 site as well. The crystal structural stability of $Na_2Ni_2TeO_6$ has been established by neutron diffraction study over 300-725 K. Nevertheless, an anisotropic thermal expansion has been found with ~3 times higher thermal expansion of the *c* axis than that of the *a/b* axes which leads to a larger separation of the Ni/TeO$_6$ polyhedral sheets. The larger separation of such layers with increasing temperature expedites higher Na-ion conductivity. Our ionic conduction study indeed reveals an enhancement of conductivity by more than two orders of magnitude with a change in temperature from 323 K to 423 K. Temperature dependent conductivity data reveal the Arrhenius behavior with an average activation energy of ~ 0.58(3) eV above 383 K. High room temperature sodium ionic conductivity and good thermal stability along with the reported reversible Na extraction/insertion process, high redox potential and reasonable capacity of 100 mAh g$^{-1}$ (as mentioned in the Introduction) make $Na_2Ni_2TaO_6$ particularly attractive for cathode material for a rechargeable Na-ion battery. The present study provides an experimental realization of Na-ion conductions in the promising battery material $Na_2Ni_2TeO_6$ and facilitates the understanding of the microscopic mechanism for Na$^+$-conductions. The present approach could be also extended to investigate other materials for Na ion conduction.

**Acknowledgment**

The authors acknowledge the help provided by A. B. Shinde for the high-temperature neutron diffraction experiments.

Table 1. The Rietveld refined lattice constants, fractional atomic coordinates, and isotropic thermal parameters [$10^2 \times B_{iso}$ (Å$^2$)] for Na$_2$Ni$_2$TeO$_6$ at 300 and 725 K. *Occ.* stands for site occupancy.

|  | Site | 300 K | 725 K |
|---|---|---|---|
| $a$(Å) |  | 5.20924(2) | 5.23801(3) |
| $c$(Å) |  | 11.15451(7) | 11.33742(9) |
| Ni | 4$d$ (2/3,1/3,0) |  |  |
| $10^2 \times B_{iso}$ (Å$^2$) |  | 0.28(2) | 1.30(5) |
| *Occ.* |  | 0.97(3) | 0.97(4) |
| Te | 2$b$ (0,0,0) |  |  |
| $10^2 \times B_{iso}$ (Å$^2$) |  | 0.19(3) | 0.93(6) |
| *Occ.* |  | 1.00(1) | 1.00(1) |
| O | 12$k$ ($x,x,z$) |  |  |
| $x/a$ |  | 0.6870(3) | 0.6777(9) |
| $z/c$ |  | 0.5942(1) | 0.5933(2) |
| $10^2 \times B_{iso}$ (Å$^2$) |  | 0.61(3) | 1.66(5) |
| *Occ.* |  | 0.94(2) | 0.95(3) |
| Na1 | 6$g$ ($x$,0,1/4) |  |  |
| $x/a$ |  | 0.3825(2) | 0.3433(4) |
| $10^2 \times B_{iso}$ (Å$^2$) |  | 1.22(2) | 4.26(3) |
| *Occ.* |  | 0.37(1) | 0.38(2) |
| Na2 | 4$c$ (1/3,2/3,1/4) |  |  |
| $10^2 \times B_{iso}$ (Å$^2$) |  | 1.22(2) | 4.26(3) |
| *Occ.* |  | 0.18(1) | 0.19(2) |
| Na3 | 2$a$ (0,0,1/4) |  |  |
| $10^2 \times B_{iso}$ (Å$^2$) |  | 1.22(2) | 4.26(3) |
| *Occ.* |  | 0.11(1) | 0.13(2) |

**Figures:**

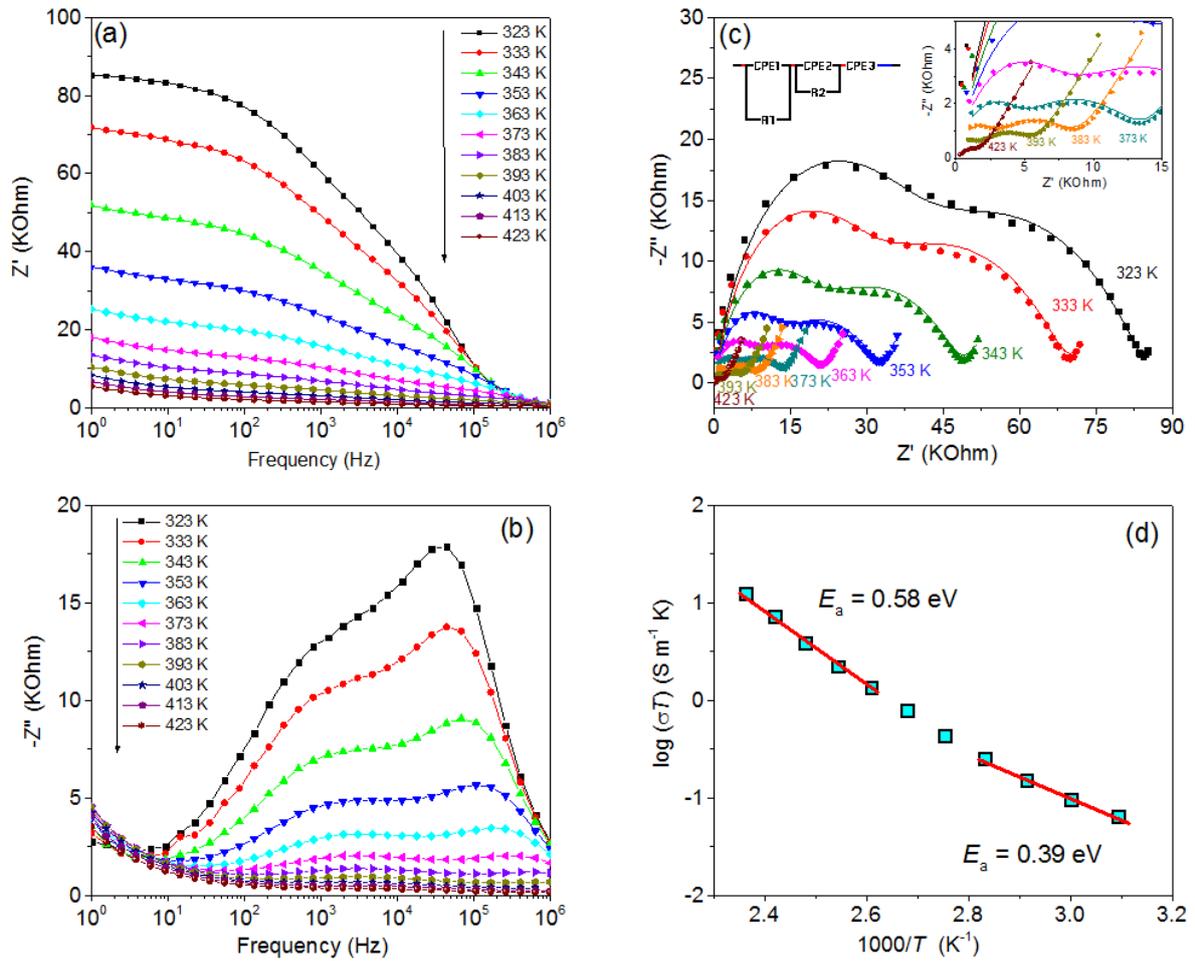

Figure 1: (Color online). The (a) real (Z′) and (b) imaginary (-Z″) parts of the impedance of $Na_2Ni_2TeO_6$ as a function of frequency ($f$) at selected temperatures over 323-423 K. (c) The Cole-Cole plot of complex impedance. The experimental data are shown by symbols. The lines show the fitted curves by an equivalent circuit (shown in the left inset). Right inset highlights the Cole-Cole plots of the high-temperature data. (d) The Arrhenius plot (square symbols) for grain conductivity (square symbols) of $Na_2Ni_2TeO_6$ derived from (c). The solid red lines through the data points are the linear fits.

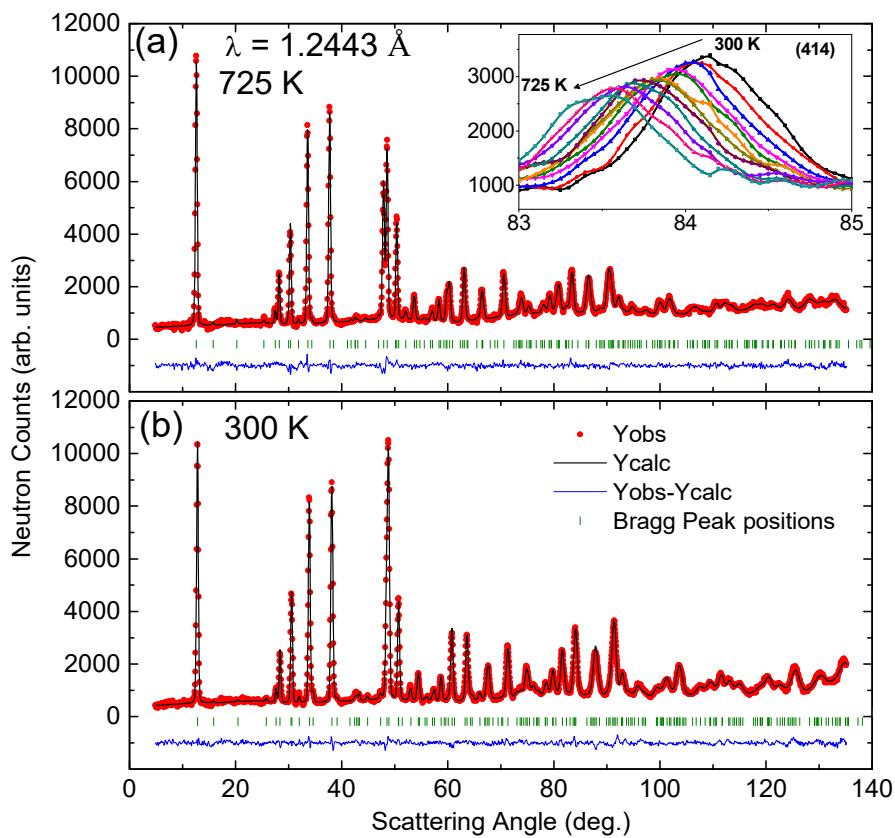

Figure 2: (Color online). Rietveld refined neutron powder diffraction patterns for Na$_2$Ni$_2$TeO$_6$ at (a) 725 K and (b) 300 K, respectively. The observed and calculated patterns are shown by filled circles and solid black lines, respectively. The difference between observed and calculated patterns is shown by the thin line at the bottom of each panel. The vertical bars are the allowed Bragg peak positions. Inset shows the temperature dependence of a representative Bragg peak (414), revealing the effect of the thermal displacement parameter on the intensity and width of the peak. With increasing temperature, a broadening of the peaks and a decline in the peak intensity are evident.

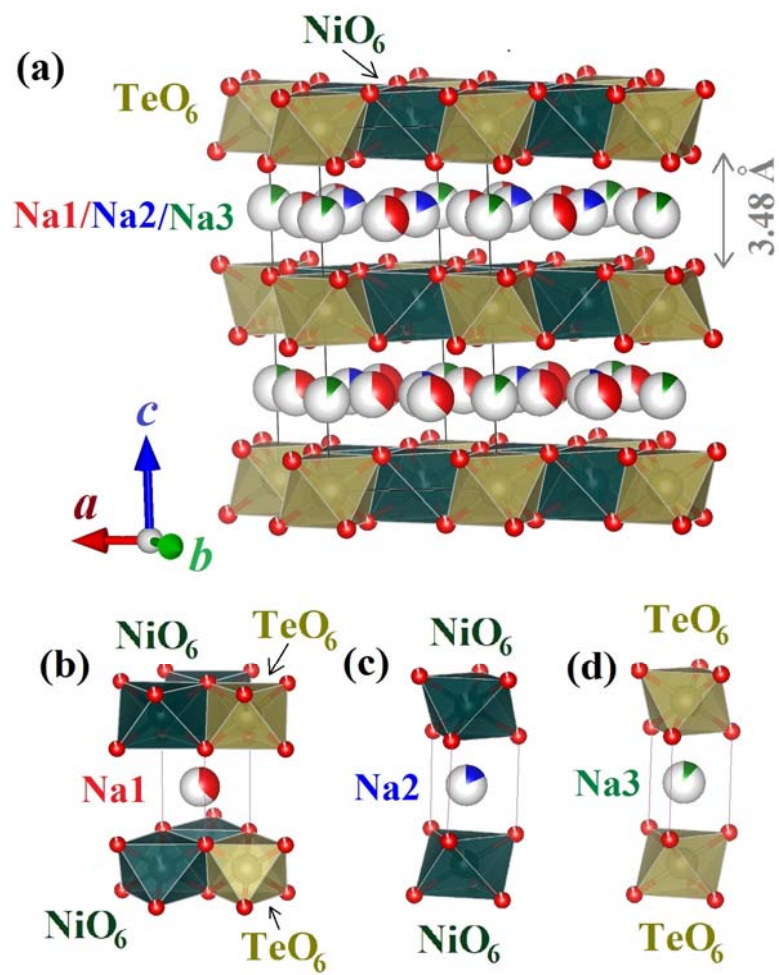

Figure 3: (Color online). (a) Layered type crystal structure of $Na_2Ni_2TeO_6$. The $NiO_6$ and $TeO_6$ octahedra are shown by greenish (dark) and yellowish (light) color. Atoms at three crystallographically different Na-sites are shown by red (Na1), blue (Na2) and green (Na3) color, respectively. The colored and white portions of a ball represent the occupied and unoccupied fraction, respectively, for a given site. The dimension of the unit cell is shown by the black thin lines. (b)-(d) Polyhedral environments of the three Na sites.

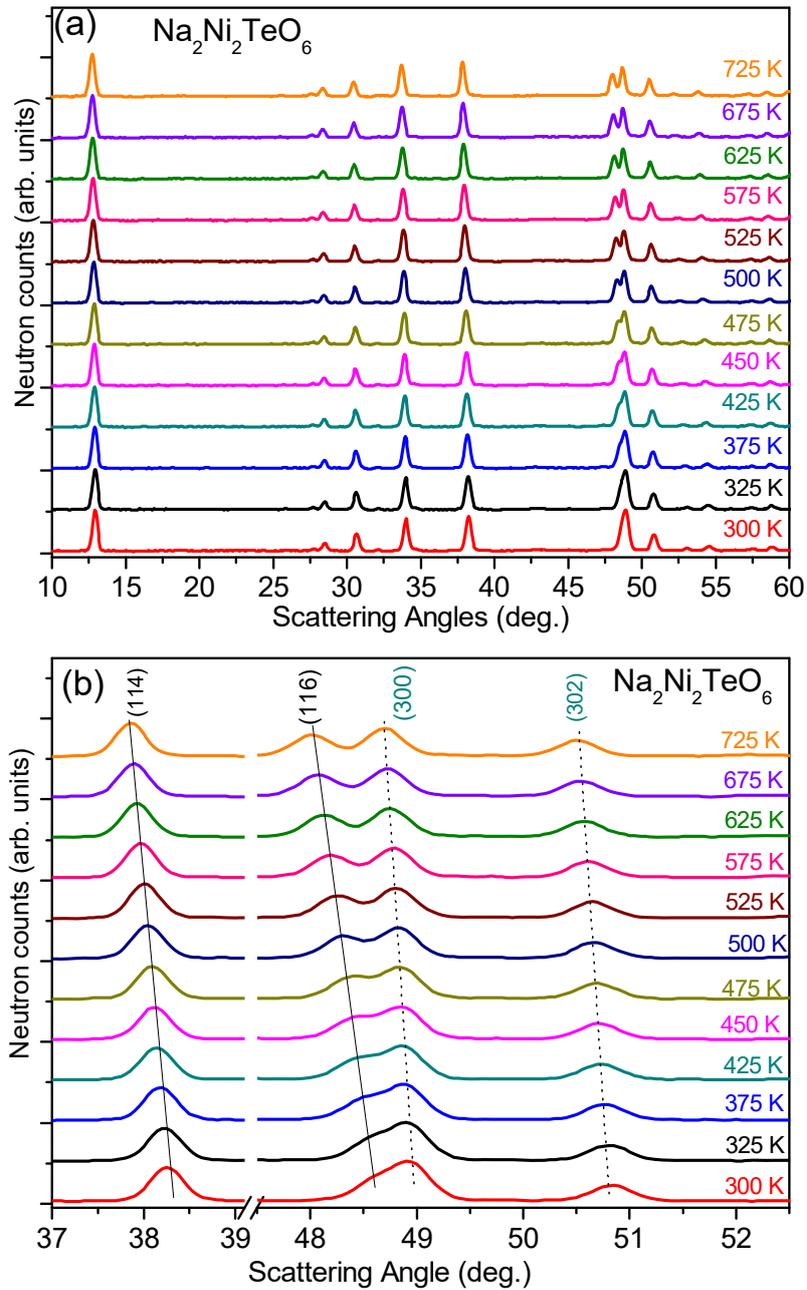

Figure 4: (Color online). (a) The temperature evolution of neutron powder diffraction patterns for Na$_2$Ni$_2$TeO$_6$ over a selected scattering angular range 10-60 deg. (b) The temperature dependences of selected Bragg peaks (114), (116), (300) and (302). The lines are guides to the eye.

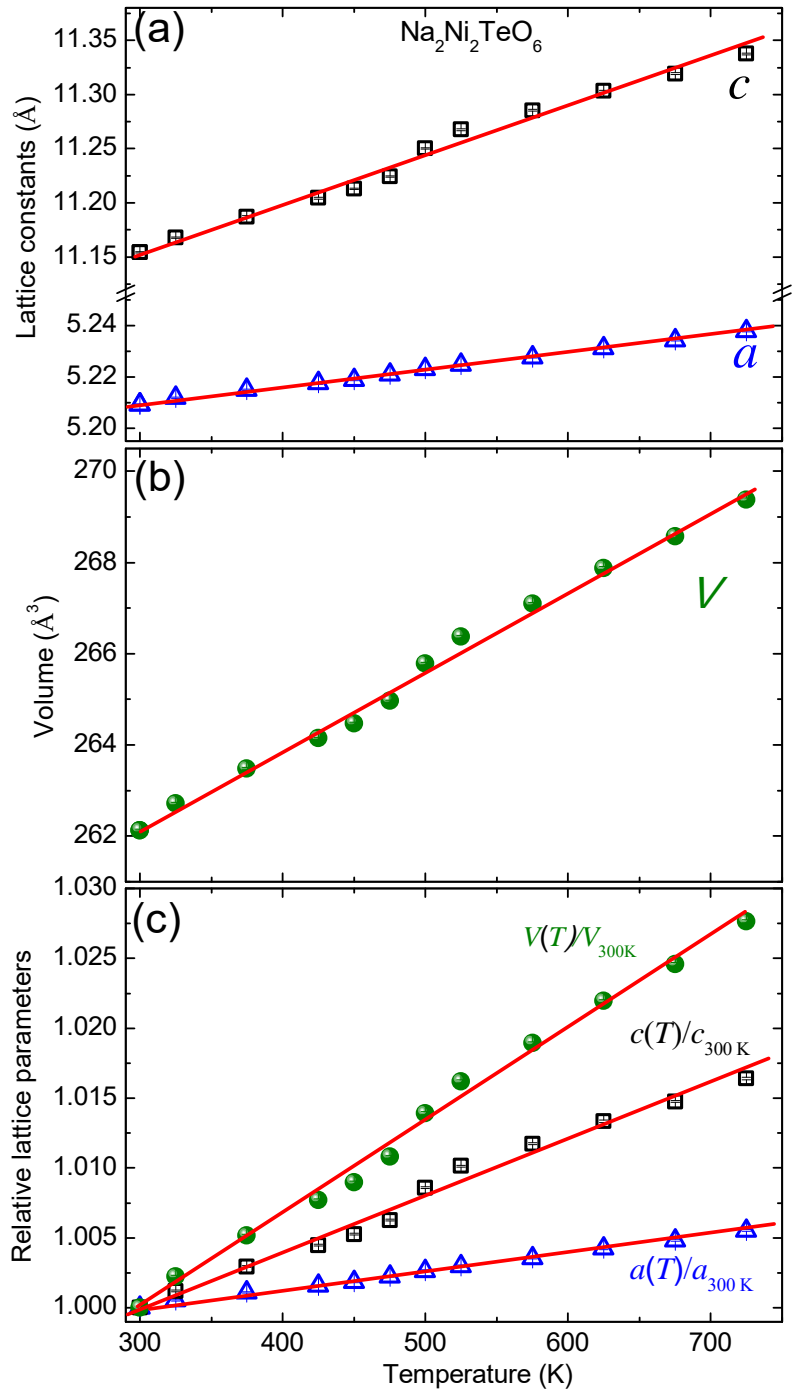

Figure 5: (Color online). The temperature dependent (a) lattice constants (*a* and *c*) and (b) unit cell volume of Na$_2$Ni$_2$TeO$_6$ over the temperature range 300-725 K. (b) The relative changes of lattice constants and unit cell volume with respect to those at 300 K. The lines are linear fits to the data.

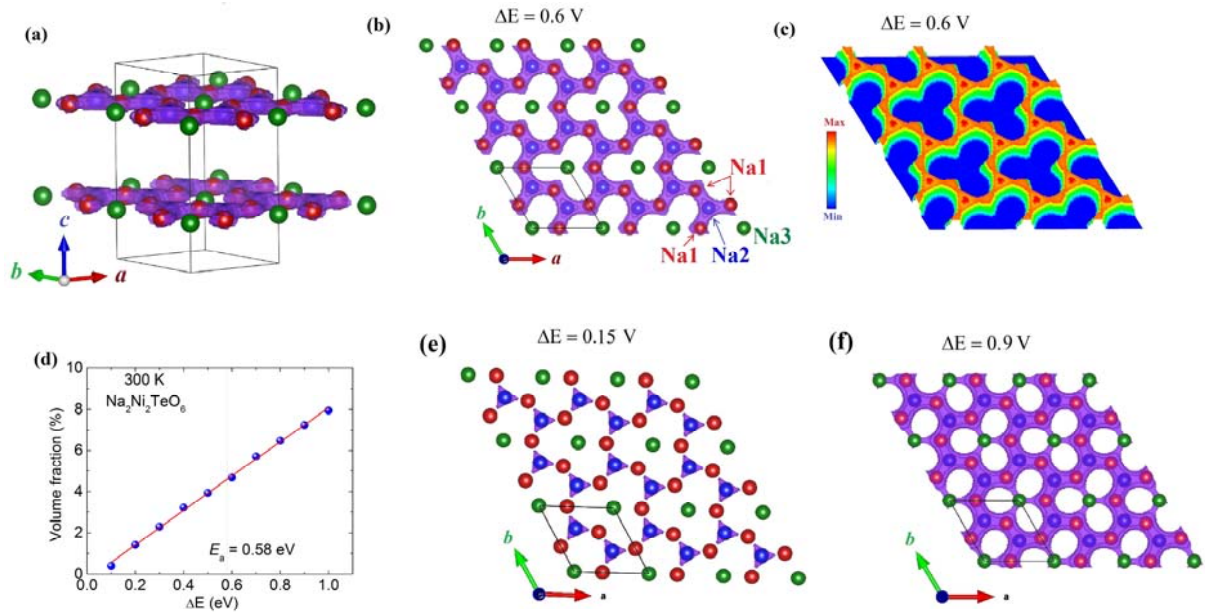

Figure 6: (Color online). (a): The bond valence energy landscape (BVEL) map (conduction pathways) of $Na_2Ni_2TeO_6$ at 300 K, represented by the shaded area. The iso-surface are drawn for the valence energy value of $\Delta E$ = 0.6 eV ($E_a$=48 eV) for Na. Two-dimensional Na-conduction within the Na-layers parallel to the *ab* plane is evident from the shaded region. For clarity, all the Na-ion sites are shown by filled balls and other atoms (Ni, Te, are O) are omitted. (b): The projection of the BVEL map in the *ab* plane (situated at *z* = 0.25) depicting the 2D circular Ni-ion conduction pattern. (c) The birds-eye view of the Na-conduction pathways in the *ab* plane (situated at z = 0.25). The color bar represents the depth profile of the conduction pathways. (d) The volume fraction for Na-ion mobility within a unit cell of $Na_2Ni_2TeO_6$ as a function of valence energy $\Delta E$ at 300 K. The data points were derived from the soft-BVS calculations of the neutron diffraction pattern at 300 K and the line is a linear fit to the data points. (e) and (f) The BVEL map in the *ab* plane (situated at *z* = 0.25) for the valence energy value of $\Delta E$ = 0.15 and 0.9 eV, respectively.

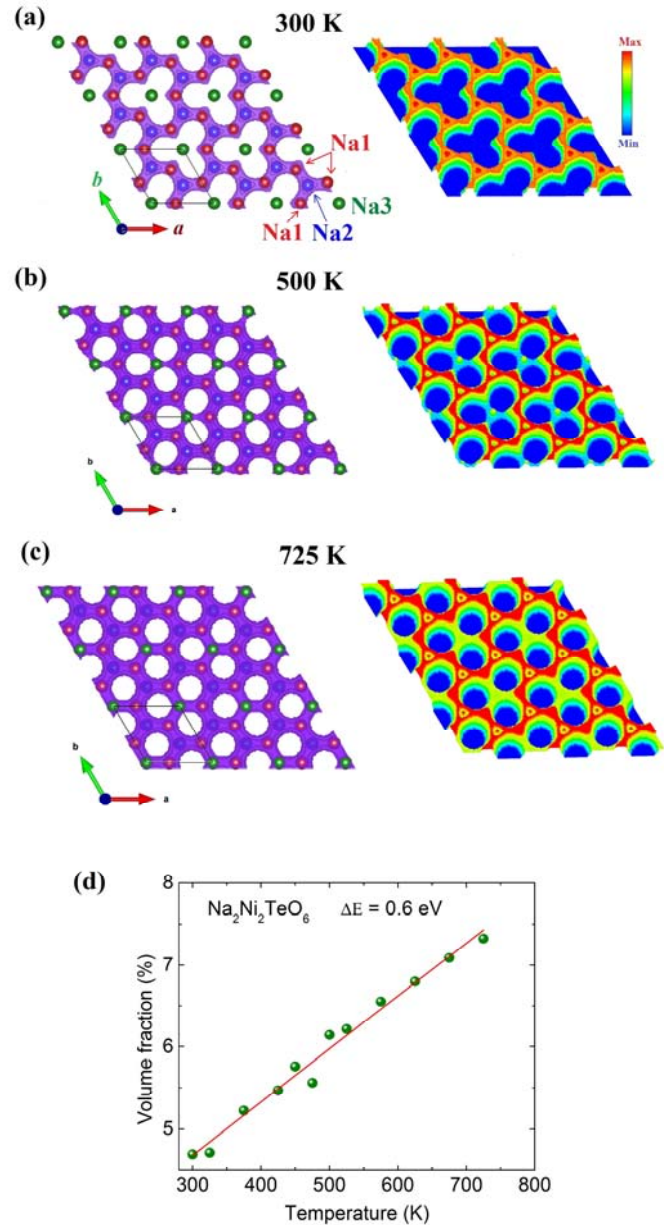

Figure 7: (Color online). (a)-(c) left: Bond valence energy landscape (BVEL) map ($\Delta E$ = 0.6 eV) of $Na_2Ni_2TeO_6$ at 300, 500 and 725 K. right: The birds-eye view of the Na-ion conduction pathways (BVEL map) and the color bar represents the depth profile of the conduction pathways. The enhanced contribution of the Na3 site to the conduction pathways is evident with increasing temperature. (d) The temperature evaluation of the volume fraction of the unit cell for Na-ion mobility of $Na_2Ni_2TeO_6$ calculated for the $\Delta E$ = 0.6 eV. The data points are derived from the soft-BVS calculations of the temperature dependent neutron diffraction patterns, and the line is the linear fits to the data.

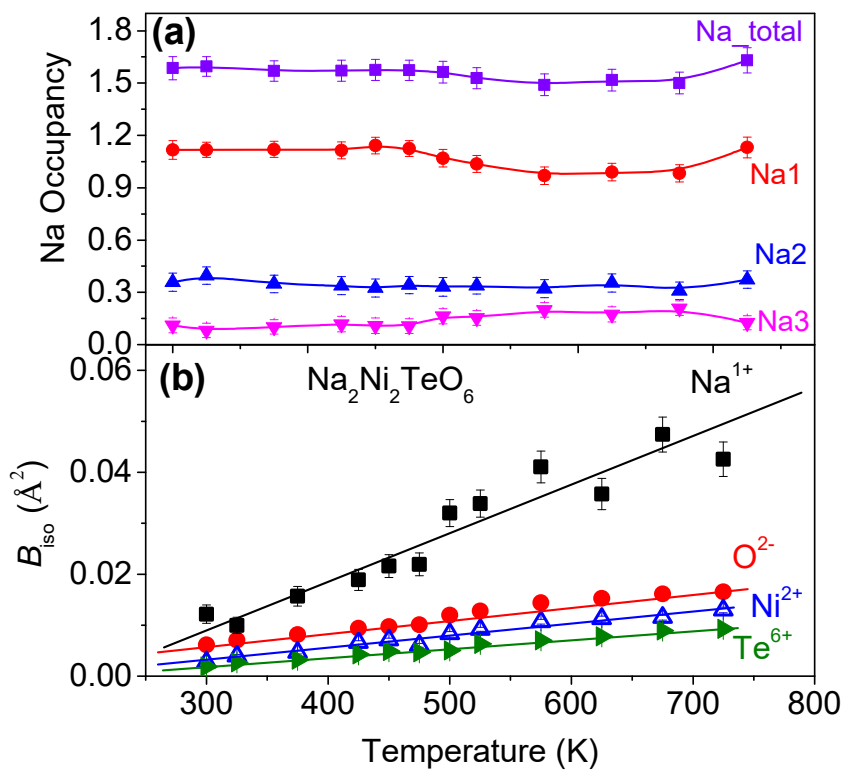

Figure 8: (Color online). Temperature dependence of the (a) Na- site occupancies and (b) isotropic thermal parameter $B_{iso}$ for the atomic sites of $Na_2Ni_2TeO_6$.

**TOC graphic**

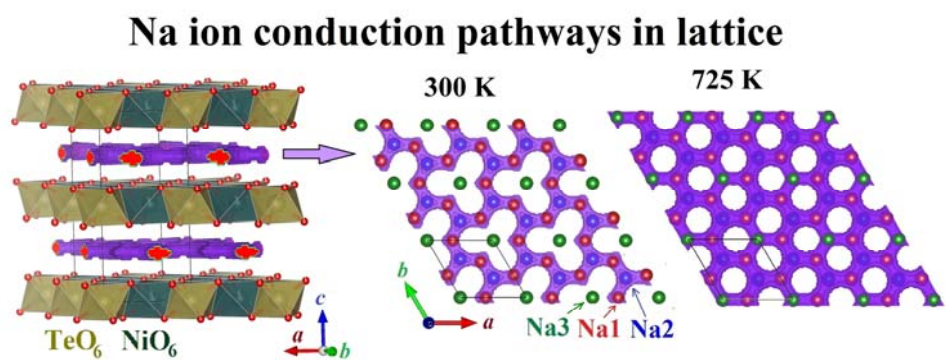

A. K. Bera *et al.*